\documentclass[aps,prd,reprint,nofootinbib]{revtex4-2}
\usepackage{graphicx}  
\usepackage{amsmath,amssymb} 
\usepackage{mathtools} 
\usepackage{physics}  
\usepackage{siunitx} 
\usepackage[colorlinks=true,linkcolor=blue,citecolor=red,urlcolor=blue]{hyperref}
\usepackage{float}  
\usepackage{subfig}  
\usepackage{caption}
\usepackage{subcaption}
\usepackage{tikz}
\usepackage{bbold}
\usepackage{orcidlink}
\usetikzlibrary{decorations.pathreplacing}

\usepackage{xcolor}

\graphicspath{{Images/}}

 \numberwithin{equation}{section}
\renewcommand{\theequation}{\arabic{equation}}
\DeclareMathOperator\arctanh{arctanh}

\makeatletter
\def\@fnsymbol#1{\ensuremath{\ifcase#1\or *\or \dagger\or \ddagger\or
   \mathsection\or \mathparagraph\or \|\or **\or \dagger\dagger
   \or \ddagger\ddagger \else\@ctrerr\fi}}

\makeatother

\begin{document}

\clearpage
\onecolumngrid 
\thispagestyle{empty}
\title{
Phase evolution of superposition target states in adiabatic population transfer}

\author{Eli Morhayim\orcidlink{0000-0002-6671-3332}$^*\dagger$}
%\thanks{These authors contributed equally to this work.}
%\altaffiliation{Current address: University of Somewhere, City, Country.}
\affiliation{Centre for Cold Matter, Blackett Laboratory, Imperial College London,  Prince Consort Road, London SW7 2AZ, United Kingdom.}

\author{Michael T. Ziemba\orcidlink{0009-0006-0691-1464}$^*$}
%\thanks{These authors contributed equally to this work.}
%\email{test@imperial.ac.uk}
\affiliation{Centre for Cold Matter, Blackett Laboratory, Imperial College London,  Prince Consort Road, London SW7 2AZ, United Kingdom.}
%\author{ Eli Morhayim\orcidlink{0000-0002-6671-3332}}\thanks{These authors contributed equally to this work.}\altaffiliation{asdf}
%\author{Michael T. Ziemba\orcidlink{0009-0006-0691-1464}}
%\thanks{These authors contributed equally to this work.}
\author{J. Lim\orcidlink{0000-0002-1803-4642}$^1$}
%\email{j.lim@imperial.ac.uk}
\affiliation{Centre for Cold Matter, Blackett Laboratory, Imperial College London,  Prince Consort Road, London SW7 2AZ, United Kingdom.}

\author{B. E. Sauer\orcidlink{0000-0002-3286-4853}$^2$}
%\email{ben.sauer@imperial.ac.uk}
\affiliation{Centre for Cold Matter, Blackett Laboratory, Imperial College London,  Prince Consort Road, London SW7 2AZ, United Kingdom.}
\date{\today}
% --- Manual Footnote Block ---
\insert\footins{\noindent \footnotesize $^*$ These authors contributed equally to this work.}
\insert\footins{\noindent \footnotesize $^\dagger$ Present address: Corpus Christi College, \\ \phantom{$^\dagger$ }Oxford, OX1 4JF, United Kingdom.}
\insert\footins{\noindent \footnotesize $^1$ Email: j.lim@imperial.ac.uk}
\insert\footins{\noindent \footnotesize $^2$ Email: ben.sauer@imperial.ac.uk}
% -----------------------------
\begin{abstract}
We consider stimulated Raman adiabatic passage (STIRAP) when the final state is a superposition of two non-degenerate states. The system consists of four states coupled by two light fields. We find the relative phase of the final superposition depends on relative amplitude, width and timing of the adiabatic transfer pulses. We discuss these results in the context of experiments measuring symmetry violation in atomic and molecular systems. 

\end{abstract}

\maketitle

\vfill 
\newpage  
\twocolumngrid  
\clearpage
\begingroup
\let\clearpage\relax
\let\clearpage\relax
\section{Introduction}

%\textcolor{blue}{Coherent control of quantum states can be achieved through various adiabatic passage techniques, including Rapid Adiabatic Passage (RAP) for two-level systems, Stark-Chirped Rapid Adiabatic Passage (SCRAP), which utilizes the AC Stark effect to control the transition dynamics, and Stimulated Raman Adiabatic Passage (STIRAP) for three-level systems using a counter-intuitive pulse sequence.}
Stimulated Raman adiabatic passage (STIRAP) is a well established technique \cite{Panda2016, Gaubatz1990} which is frequently used for efficient state transfer while being robust to experimental imperfections. In some experiments, the only quantity of interest is the probability of populating the target state, and here STIRAP is often superior to other methods. In other experiments, the desired final state is a superposition of two quantum states whose relative phase needs to be controlled carefully. This is especially important in atomic and molecular physics experiments that test fundamental symmetries, such as measurements of the electron's electric dipole moment \cite{Athanasakis2025,Andreev2018,Alauze2021} or of parity violation and time-reversal-symmetry violation in the nucleus \cite{Altuntas2018}. In such experiments, the quantity to be measured is encoded in this phase difference, so it is important to know the exact phase produced by STIRAP and its robustness to experimental imperfections. Here, we consider the phase difference produced by STIRAP when the target state is a superposition of two non-degenerate states.

Many extensions to STIRAP have already been considered. These include the effects of background states \cite{Jakubetz2012}, transfer through chains of intermediate states \cite{Marte1991}, or parallel linkages with differing intermediate states \cite{Vitanov1999}. In the `tripod STIRAP' configuration the typical two-ground plus one-excited STIRAP scheme is extended with a third ground state linked by an additional light field. Any particular superposition of two target states can be prepared by choosing the appropriate pulse ordering and overlap of the light fields \cite{Vitanov2016, Unanyan1998,Theuer1999}. In this paper we consider the situation in which both target states are addressed by a \textit{single} light field. We restrict our treatment to the case where the two-photon detuning (the frequency difference between the transition being driven and the difference of the two light fields) is small and symmetric for the two target states, as shown in figure \ref{fig:figure1a}.  Here, the STIRAP  evolution is not through a single dark state but instead through a superposition of two quasi-dark eigenstates. We find that the phase in the final target superposition first reaches a plateau in the early stages of STIRAP before becoming the linear phase evolution expected for two states with a constant energy difference.

\section{The Model}\label{sec:2}
Figure \ref{fig:figure1a} illustrates our model, which consists of four quantum states $\ket{g}$, $\ket{e}$, $\ket{\uparrow}$ and $\ket{\downarrow}$ linked by `pump' and `Stokes' light. In this paper we restrict our treatment to a two-photon detuning $\delta$ that is equal and opposite for the two target states. The single photon detuning is $\Delta$. 
We derive the Hamiltonian by applying the rotating wave approximation and transferring to a rotating frame in which the Hamiltonian assumes a form which is time-independent apart from the slow variation of the Rabi amplitudes which drives the adiabatic evolution. With the pump and Stokes Rabi rates $\Omega_p$ and $\Omega_s$ this yields (with $\hbar=1$)
\begin{align}
    & \hat{H} = \frac{1}{2}\begin{pmatrix}
        0& \Omega_p & 0 & 0 \\
        \Omega_p &2 \Delta & \Omega_s & \Omega_s \\
        0 & \Omega_s & 2\delta& 0 \\
        0 & \Omega_s & 0 &-2 \delta\\
    \end{pmatrix}. 
    \label{hamiltonian}
\end{align}

In canonical three-level STIRAP with a single target state, adiabatic state transfer is achieved through a dark state composed of the initial and final states only. The content of this dark state depends on the mixing angle $\arctan (\Omega_p/\Omega_s)$. With `counter-intuitive' ordering
\begin{figure}[bh]
    \includegraphics[scale=0.55]{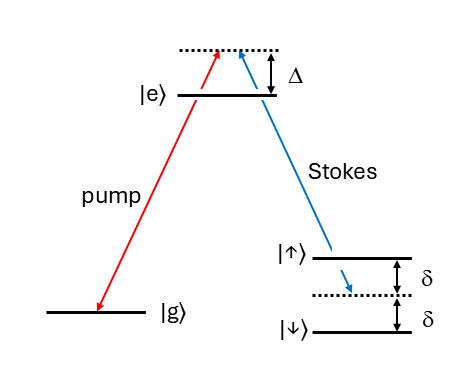}
    \caption{Energy levels and coupling in the four-sate system. The splitting $\delta$ is very small compared to $\Delta$ and the $\ket{e}-\ket{g}$ energy difference. }
    \label{fig:figure1a}
\end{figure}
(Stokes light leading pump light), the dark state evolves from the initial to the final state. 
In this work, we will consider gaussian profiles for the $[p,s]$ pulse envelopes:
\begin{equation}
    \Omega_{[p,s]} (t) = \Omega_{0[p,s]}\exp{-\frac{(t-\mu_{[p,s]})^2}{2T^2}}.
    \label{rabifreqdetails}
\end{equation}
In common with three-level STIRAP, there is a strong dependence of the success of population transfer on $\delta$ with four levels. In the simulations, population transfer efficiency decreases to $99\%$ 
 for $\delta \sim 20\, T^{-1}$ (with peak Rabi frequencies of $248\, T^{-1}$ and a pulse separation of $1.5\, T$). For larger values of $\delta$, it deteriorates further. However, for most relevant applications and the scenario of interest in this paper the two photon detuning is much smaller than these values such that the evolution is sufficiently adiabatic.
 
  In our four-level system, when $\delta=0$, the initial state evolves into an equal superposition of the final states with zero relative phase \cite{Unanyan1998}. However, the phase evolution is non-trivial in the presence of a small $\delta$, as seen in fig. \ref{fig:numericalPhaseTwoPhoton}. 
\begin{figure}
    \includegraphics[scale=0.65]{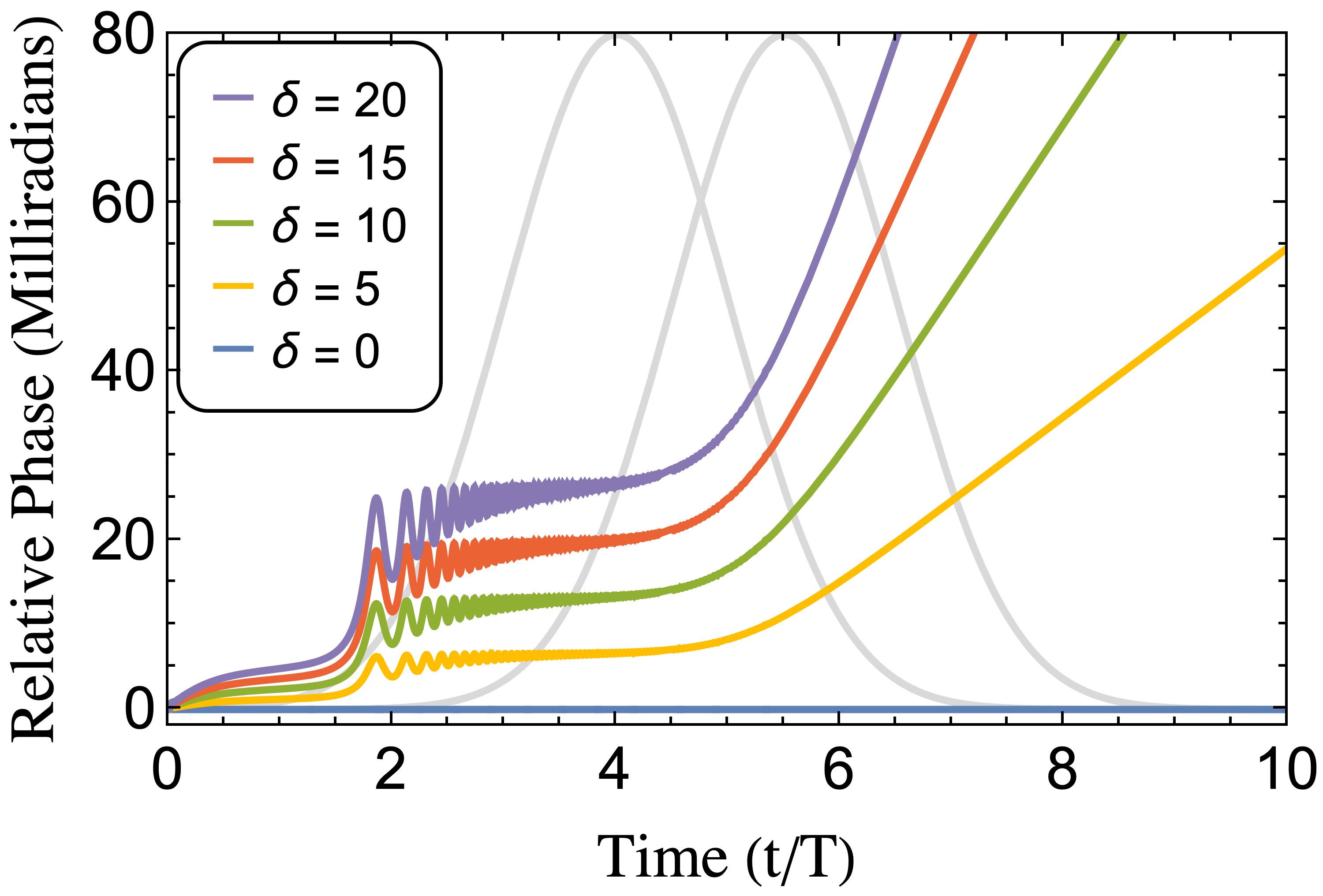}
    \caption{Numeric calculation of the time evolution of the relative phase between states  $\ket{\uparrow}$ and $\ket{\downarrow}$ for a range of different $\delta$ (in units of $10^{-3} T^{-1}$). The envelopes of the Stokes and pump light are shown in gray. The relative phase evolves in two steps. First, a sudden jump to a plateau appears before a switch to linear phase accumulation 2$\delta t$. }
\label{fig:numericalPhaseTwoPhoton}
\end{figure}
For non-zero $\delta$, our numeric results show that, prior to significant state transfer, the relative phase between states $\ket{\uparrow}$ and $\ket{\downarrow}$ undergoes fast oscillations before settling to a plateau. (The time $t=0$ is chosen to be before significant population transfer occurs, but late enough that the Rabi frequencies are much larger than $\delta$.) Our analysis suggests that these oscillations occur due to non-adiabatic leakage of population into the eigenstates $\ket{c}$ and $\ket{d}$ (see fig. \ref{fig:4lvlEigenstateEvolution}), which do not initially overlap with the ground state and therefore should not contribute to the population transfer in STIRAP.  As the state transfer finishes, the phase settles into the linear evolution expected for two states with finite energy difference.  

\section{Analytic Treatment of the Phase Evolution}

 An analytic treatment of STIRAP with a degenerate superposition of final states ($\delta = 0$) has been considered in ref. \cite{Theuer1999}. Such a system starts in one of two dark states. Non-adiabatic interactions with the other dark state determine which final superposition is prepared. 

For the $\delta \neq 0$ case considered here, the main features of the phase evolution shown in fig. \ref{fig:numericalPhaseTwoPhoton} can also be explained by an analytic treatment. We set $\Delta = 0$ for simplicity (see  appendix \ref{sec:AppendixRelativePhase} for a perturbative treatment) and assume that the process is fully adiabatic such that the initial (field-free) state follows the adiabatic evolution of the eigenstates of the full Hamiltonian of eq. \ref{hamiltonian}. 
The two eigenstates which overlap with the initial state at the start and final state at the end are
\begin{widetext}
\begin{gather}
    \ket{a}=N\left[\frac{\Omega_p}{2E_a}\ket{g} +\ket{e} + \frac{1}{E_a - \delta}\frac{\Omega_s}{2}\ket{\uparrow}+\frac{1}{E_a +\delta}\frac{\Omega_s}{2}\ket{\downarrow} \right], \label{eq:stateA}\\
    \ket{b}=-N\left[\frac{\Omega_p}{2E_a}\ket{g}-\ket{e}+\frac{1}{E_a + \delta}\frac{\Omega_s}{2}\ket{\uparrow}+\frac{1}{E_a -\delta}\frac{\Omega_s}{2}\ket{\downarrow} \right],
    \label{eq:stateB} \\
    |N|^2 = \left[ 1 + \frac{\Omega_p^2}{4E_a^2} + \frac{\Omega_s^2}{4(E_a-\delta)^2} + \frac{\Omega_s^2}{4(E_a +\delta)^2}\right]^{-1}, \\
        E_a = -E_b = \frac{-1}{2\sqrt2}\sqrt{4\delta^2+\Omega_p^2 + 2\Omega_s^2-\sqrt{-16\delta^2\Omega_p^2+(4\delta^2+\Omega_p^2+2\Omega_s^2)^2}}.\label{eq:ABEigenengergies}
\end{gather}
\end{widetext}
The other two eigenstates do not participate in the STIRAP population transfer and so need not be considered here. With $\delta\neq 0$, the initial state is a superposition of the two nearly degenerate quasi-dark states  $\ket{a}$ and $\ket{b}$ (see appendix \ref{sec:AppendixEigenstates}). There is no coupling as $t\rightarrow -\infty$. With $\Omega_s=\Omega_p=0$  the eigenstates in eqns. \ref{eq:stateA} and \ref{eq:stateB} reduce to 
\begin{align}
    \ket{a(-\infty)} = \frac{1}{\sqrt{2}}(\ket{e} + \ket{g}), \\
    \ket{b(-\infty)} = \frac{1}{\sqrt{2}}(\ket{e} - \ket{g}).
\end{align}
The initial state is therefore
\begin{equation}\label{InitialState}
    \ket{\psi(-\infty)} =  \ket{g} = \frac{1}{\sqrt{2}}\left(\ket{a(-\infty)} - \ket{b(-\infty)}\right).
\end{equation}
Although $\ket{a}$, $\ket{b}$ individually contain $\ket{e}$, this superposition is dark for $\delta = 0$.

Numeric calculation shows no geometric phase evolves for $\ket{\Psi(t)}$, which is expected to be zero (see  appendix \ref{sec:AppendixGeometricPhase}). 
The adiabatic evolution of the state is thus
\begin{align}
    \ket{\psi(t)} = \nonumber  \frac{1}{\sqrt{2}}&\left(e^{-i\int_{-\infty}^t E_a(t') \ dt' \,}\ket{a(t)}\right.\\ &\left. -e^{-i\int_{-\infty}^t E_b(t') \ dt' \,}\ket{b(t)}\right).
    \label{protostate}
\end{align}
 This state is a superposition of two eigenstates which merely acquire a relative phase due to their energy difference. Fig. \ref{fig:4lvlEigenstateEvolution} shows an example of the time dependence of the composition of the eigenstates as given by eqns. \ref{eq:stateA}, \ref{eq:stateB}  as well as their eigenenergies according to eq. \ref{eq:ABEigenengergies}. The state at early times evolves into a superposition of the target states $\ket{\uparrow}$ and $\ket{\downarrow}$.

\begin{figure*}
    \centering
    \includegraphics[scale=0.45]{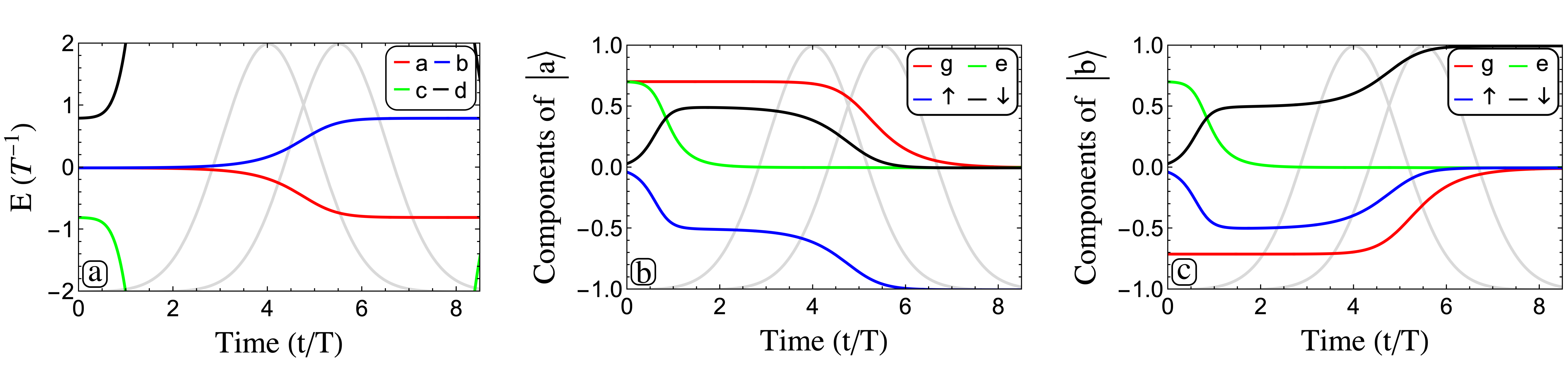}
    \caption{Time evolution of the eigenenergies (a) and components of the two eigenstates $\ket{a}$ (b) and $\ket{b}$ (c). Both states start as a superposition of initial and excited state and then gradually evolve into one of the target states. A  fast reorientation from and back to the initial composition occurs as the light turns on and off.}
    \label{fig:4lvlEigenstateEvolution}
\end{figure*}

The relative phase of the final superposition of $\ket{\uparrow}$ and $\ket{\downarrow}$ is found by evaluating
\begin{equation}
    \phi(t) := \arg(\braket{\downarrow}{\psi(t)})-\arg(\braket{\uparrow}{\psi(t)}).
\label{eq:phase}
\end{equation}
The condition $\Omega_s \gg \delta$ holds in the plateau which allows for an approximation yielding a simple expression for the phase (see appendix \ref{sec:AppendixRelativePhase})
\begin{equation}
    \phi(t) = -\frac{2\delta}{E_a(t)}\int_0 ^t E_a(t') dt'. \label{int}
\end{equation}
This phase can be evaluated explicitly for gaussian envelopes of $\Omega_s$, $\Omega_p$:
\begin{equation}
    \phi(\tau) = \epsilon (\tau)\ln{\left[\left(\frac{1-\sqrt{1+\gamma}}{1+\sqrt{1+\gamma}}\right)\left(\frac{1+\sqrt{1+\gamma e^{-\tau}}}{1-\sqrt{1+\gamma e^{-\tau}}}\right)\right] \label{fig:phitau}} 
\end{equation}
with definitions $r=\Omega_{0s}/\Omega_{0p}$ and
\begin{equation}
    \tau = \frac{2(\mu_p -\mu_s)}{T^2}t, \label{75}
\end{equation}    
\begin{equation}
    \epsilon (\tau) =\frac{\delta^2T^2}{(\mu_p-\mu_s)E_a(\tau)},\label{76}
\end{equation}    
 \begin{equation}    
    \gamma = 2r^2\mathrm{exp}\left(\frac{\mu_p^2-\mu_s^2}{T^2}\right).
    \label{77}
\end{equation}

Fig. \ref{fig:numericalPhaseParameterDependence} (a) compares numeric results to the analytic phase given by eq. \ref{fig:phitau}. Apart from the early behavior and the fast oscillations in the numeric results, the calculated phases are in good agreement. In fig. \ref{fig:numericalPhaseParameterDependence} (b)--(e) we explore the dependence of the height of the plateau on the STIRAP parameters. We see that the height is independent of the peak Rabi frequencies, but linearly dependent on $\delta$ and on the width of the STIRAP pulses. 
   As is shown in (d), there is a sharp increase in the phase at small pulse separations, which tends to zero as the separation increases. The plateau height (e) grows nearly linearly with the width of the pulses. The numerical and analytic predictions diverge at low Rabi frequencies, small beam separations, and large standard deviation. These parameters are also regions where the adiabatic condition fails, with lower population transfer efficiency and increased rippling. This highlights a shortcoming of the analytical formula, which assumes the process to be sufficiently adiabatic. The formula cannot capture the oscillations at early times which are due to diabatic transitions out of the carrier eigenstates. The assumption $\Omega_s \gg \delta$ is needed to obtain a closed form solution - it fails at early times.
   
\begin{figure*}
    \centering
    \includegraphics[scale=0.58]{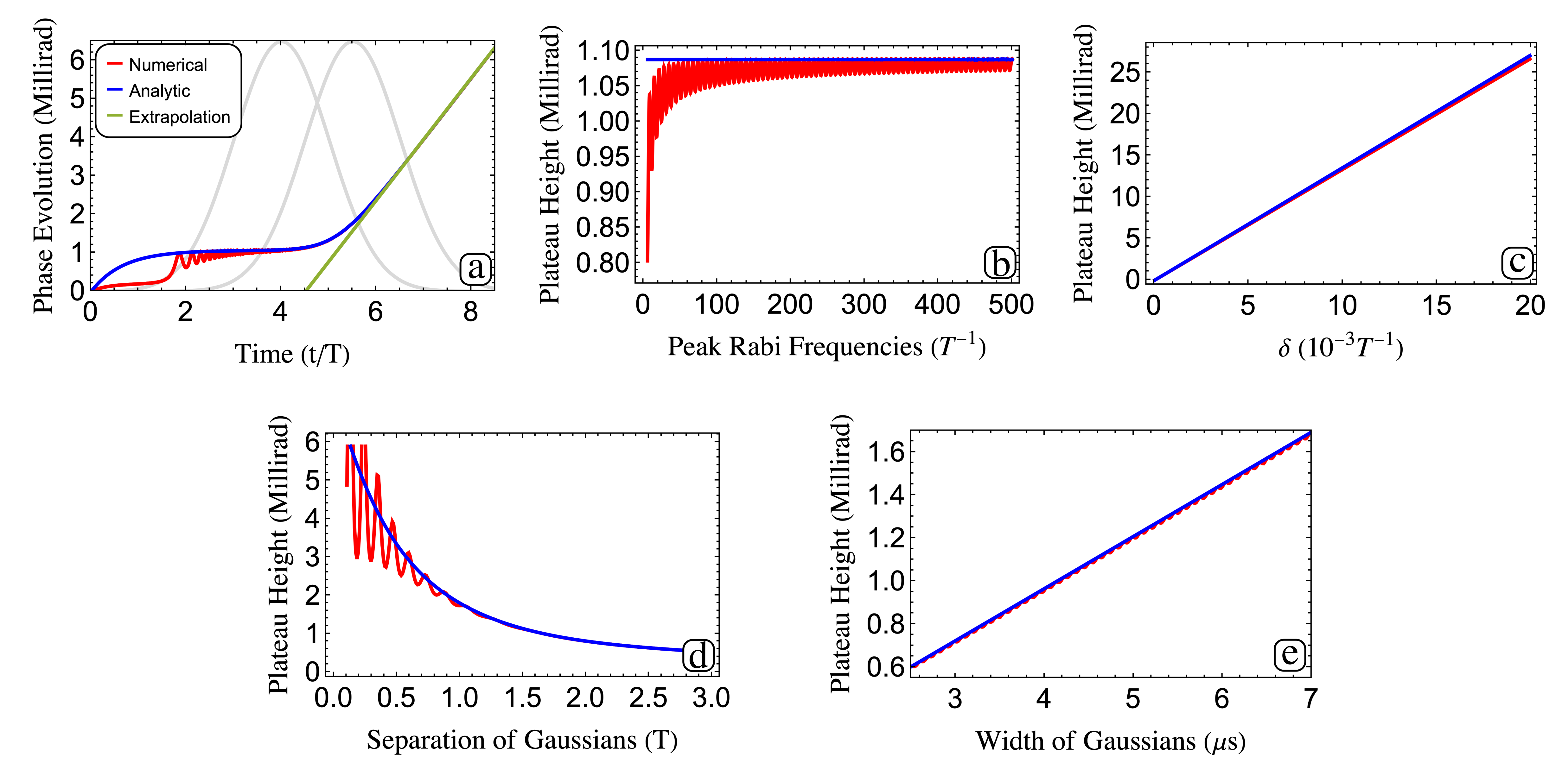}
    \caption { Comparison of numeric and analytic results for the relative phase between the two target states. (a) Time evolution of the relative phase. The analytical and numerical results are in excellent agreement except at the earliest times. (b--e) Comparison of the relative phase in the plateau region. There is little dependence on peak Rabi frequency (b) and nearly linear dependence on the detuning $\delta$ (c). The separation of the beams alone has been varied in (d) while both the width $T$ of the pulse envelopes and their separation has been varied in (e) such that the separation is always at 1.5T. All plots use a peak Rabi frequency of 248 $T^{-1}$ and a beam separation of $1.5T$ except where these are the variables.}
    \label{fig:numericalPhaseParameterDependence}
\end{figure*}
\vspace{2ex}
\subsection*{Characterization of the transition from plateau to  phase evolution}

As shown in fig. \ref{fig:numericalPhaseParameterDependence} (a), extrapolating the linear phase evolution to zero phase suggests that the effective phase evolution begins near the crossing point of the two profiles. An expression for the onset of phase evolution (the x-intercept) can be found by expanding the analytic expression for the phase (eq. \ref{fig:phitau}) at a time after the STIRAP transfer. This yields 
\begin{equation}
   \phi(t) = 2\delta (t-t_1-t_2)
   \label{eq:midphase}
\end{equation}
with        
     \begin{equation}
        t_1 = \frac{\mu_p^2-\mu_s^2+2T^2\ln(r)}{2(\mu_p-\mu_s)}, 
     \end{equation}
     \begin{equation}
        t_2 = \frac{T^2}{2(\mu_p-\mu_s)}\ln\left[\frac{1}{2}\frac{\sqrt{1+2r^2e^{\frac{\mu_p^2-\mu_s^2}{T^2}}}+1}{\sqrt{1+2r^2e^{\frac{\mu_p^2-\mu_s^2}{T^2}}}-1}\right].
        \label{eq: phase onset}
    \end{equation}
This expression for the phase contains the linear evolution expected, $2\delta t$ with two additional terms, $2\delta t_1$ and $2\delta t_2$. The time $t_1$ is the time at which the Rabi intensity profiles intersect (easily seen if we set $r=1$ so $t_1 = (\mu_p+\mu_s$)/2). The term $t_2$ is usually smaller. For example, with $r=1$, $\mu_p= 5.5 T$ and $\mu_s = 4 T$ then $t_2 \simeq -0.23 T$. This term depends on the pulse timings, the ratio of their amplitudes, and the pulse envelope width. In fig. \ref{fig:t2ParameterDependences} we compare analytic and numeric values of $t_2$ as these parameters are varied. 

\begin{figure}
    \centering
    \includegraphics[scale=0.35]{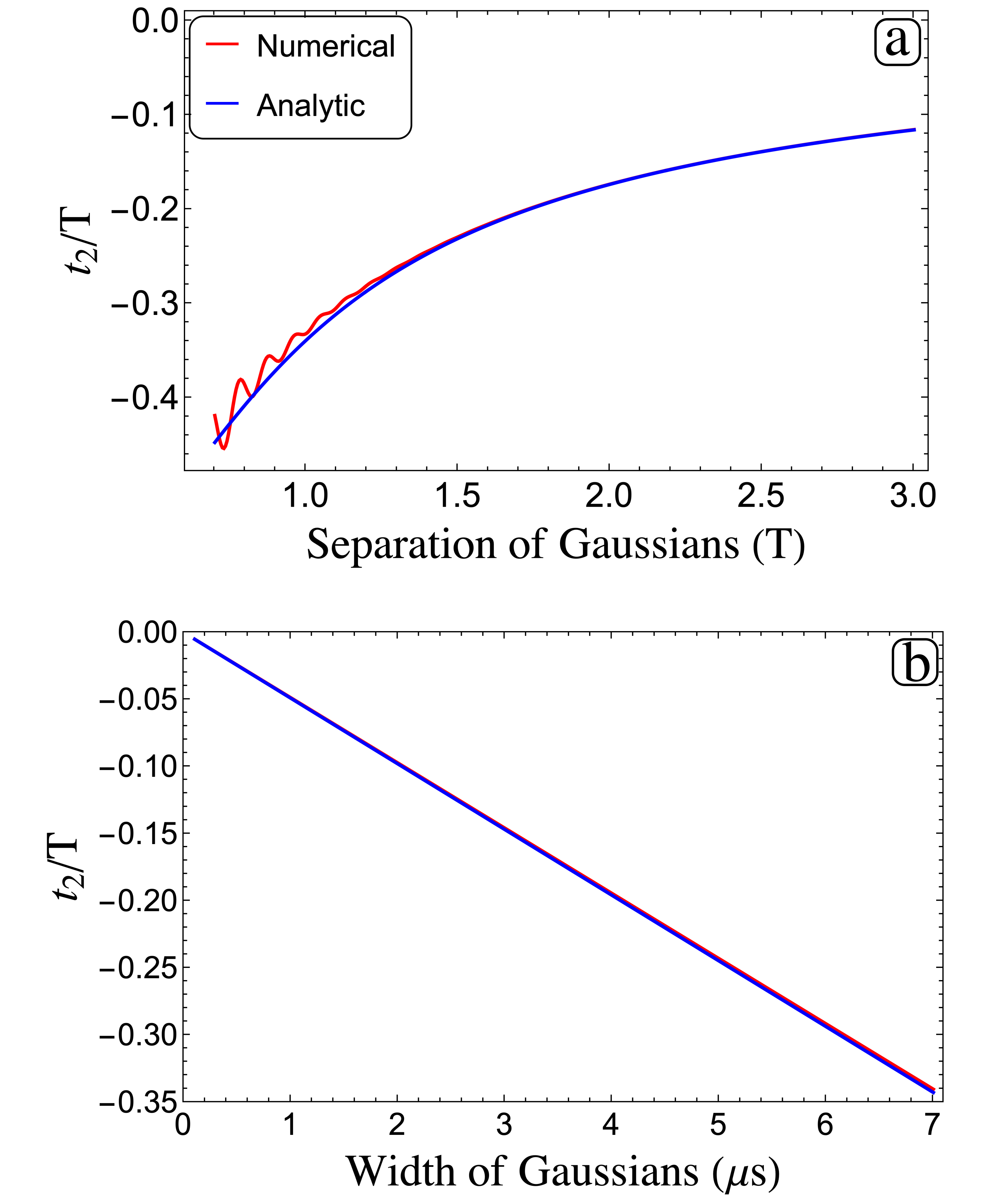}
    \caption{The dependence of the phase correction term ($t_2$) on beam parameters. Numerical simulations shown in red are compared with the analytical formula (eq. \ref{eq: phase onset}) shown in blue.  With other parameters fixed, (a) varies the separation of the beams $\mu_p-\mu_s$, and (b) the Rabi envelope width $T$.}
    \label{fig:t2ParameterDependences}
\end{figure}

\section{Discussion}
Our treatment of STIRAP with non-degenerate final states shows that, in addition to the phase accumulated due to the states' energy difference, there are phase terms which depend on the parameters of the STIRAP process itself. Here, we discuss this result in the context of an experiment to probe time reversal violation in YbF, \cite{Jenkins2026} which uses two STIRAP processes to prepare and analyze a spin superposition. The splitting of these degenerate states $\delta$ is lifted by a small applied magnetic field. The experimental signature of time reversal violation is a phase change correlated with the direction of an applied electric field. The overall precession time is much larger than  $t_1$ and  $t_2$, so the additional STIRAP phases are very small. Their scale is set by the pulse width $T$, which in this experiment is only about 0.002 of the precession time. 

An ideal experiment would have a sensitivity limited only by shot noise. In the pulsed beam YbF experiment, the number of molecules per pulse $N$ is at least $10^6$, so phase noise $\delta\phi \approx 1/\sqrt{N}= 10^{-3}$~ rad could be significant. Variation of the STIRAP laser beam positions (e.g. due to vibrations) vary the STIRAP parameters, but the phase noise that results from this effect is insignificant. 

The analytic results presented here show that the peak Rabi frequencies and their ratio $r$ have little effect on $t_1$ and $t_2$, with a less than 1\% discrepancy between numeric and analytic solutions. In the molecular beam experiment $\Delta t_1 \approx T \delta r / r$. 
It is easy to keep $r$ stable to 1\%, leading to negligible fluctuations in $t_1$. A similar analysis for $t_2$ leads to similar conclusions.

 Any phase change correlated with the direction of the applied electric field could result in a systematic effect at the level of the target sensitivity. Of particular concern is $\Delta$, since this detuning varies with the Stark shift of the optical transition, which is large \cite{Condylis2005}. The phase eq. \ref{eq:midphase} was calculated assuming $\Delta=0$. However, at least to first order, the effect of $\Delta$ is zero (see appendix), and numeric modeling of the experiment shows that variations in $\Delta$ make insignificant contributions to the phase. For example, the correction term ($t_2$) only gets a 2.8\% change when $\Delta$ is at 400 $T^{-1}$ ($\sim $ 84 MHz).
 
 None of the other relevant parameters of eq. \ref{eq:midphase} are correlated with the electric field magnitude or direction. We can speculate on higher order effects which could conspire to result in electric field correlated changes of $t_1$ or $t_2$.  
In the experiment, two STIRAP sequences are employed: the first prepares the superposition state, and the second transfers it back to the original ground state after the interaction with electric and magnetic fields. The STIRAP pulses are formed by two elongated, partially overlapping gaussian laser beams. Ideally, the beams' k-vectors, the velocity of the molecules, and the electric field are all perpendicular to one another. A slight angle between the major axes of the two ellipsoidal laser beams leads to a gradient of the beam separation along the electric field direction. If the position of the molecular beam is correlated with electric field direction (for example by changing fringe fields due to patch potentials on the electrodes), the tilt of the beam profiles generates a change in $t_2$ by changing the pulse separation. The magnitude of the possible effect is shown in fig. \ref{fig:t2ParameterDependences} (a). The phase gradient at 1.5 T separation is about 0.1. With an angle between the pulse envelopes of $1^\circ$ and a correlated displacement along the electric field of \SI{15}{\micro\meter} (which is about $10^{-4}$ the length of the laser beams), $t_{2,corr}$ is about \SI{0.2}{\nano\second}. 
For $2\delta = 2\pi\times$\SI{25}{\hertz}, the correlated change in phase is about 30$\mu$rad.  The accumulated effect for these parameters is smaller than the expected experimental sensitivity. 

The second STIRAP sequence transfers the superposition state back to the initial state to analyze its phase. It can be described within the same framework as the first. Its phase response to variations in the STIRAP parameters follows the same scaling and does not introduce additional sensitivity, a finding confirmed by numeric simulations shown in appendix \ref{sec:reversestirap}. We conclude that even though the phase of the superposition depends on the STIRAP parameters, it is unlikely that changes in the parameters are significant for the YbF  experiment.

\section{Conclusion}
In this work, we studied the evolution of the relative phase between two non-degenerate target states, in which both target states are simultaneously addressed by the same Stokes field. We derived an analytical expression for the relative phase in the fully adiabatic limit and compared it with numerical simulations. The phase evolution exhibits a characteristic structure: a rapid jump to an early-time plateau followed by a gradual crossover to the linear phase evolution expected for two states with a finite energy separation. We showed that both the plateau height and the onset of the linear regime depend on the laser beam parameters.

This study was motivated by concerns that STIRAP-based preparation and analysis of superposition states could introduce systematic phase shifts in precision measurements, such as experiments searching for the electric dipole moment of the electron. Our numerical results show that, for the range of plausible experimental parameters, these effects are small compared to the level of current experimental precision. The analytical results also show that the induced phase shifts do not scale with experimental parameters in the same way as an electron electric dipole moment signal, suggesting that they are unlikely to constitute an important  source of systematic error. The effects discussed here may become relevant for future experiments with improved sensitivity. 

\section{Acknowledgments}
This research was undertaken thanks in part to funding from the Career Development Fund at Corpus Christi College Oxford. This work has been supported by the ``Table-top experiments for fundamental physics'' program, sponsored by the Gordon and Betty Moore Foundation (Grants 8864 \& GBMF12327, DOI 10.37807/GBMF12327), Simons Foundation, Alfred P. Sloan Foundation, and John Templeton Foundation Grants (G-2019-12505 \& G-2023-21035). The research has also been supported by UKRI under Grants EP/X030180/1, ST/V00428X/1 and ST/Y509978/1. We thank M.\ R.\ Tarbutt for helpful comments.
\endgroup

\newpage

\appendix
\numberwithin{equation}{section}
\renewcommand{\theequation}{\thesection\arabic{equation}}
\renewcommand{\braket}[2]{\langle #1 |#2 \rangle}

\section{Derivation of the eigenstates}\label{sec:AppendixEigenstates}

We construct the eigenstates of the coupled system in the bare basis $\ket{g}$, $\ket{e}$, $\ket{\uparrow}$, $\ket{\downarrow}$ without explicitly diagonalizing the hamiltonian (1). We label the bare states $\ket{j}$ and the adiabatic eigenstates $\ket{i}$ with eigenenergies $E_i(t)$. With the hamiltonian acting on the adiabatic state, we have
\begin{equation}\label{eq:SupplPhase1}
    \mel{j}{\hat{H}}{i} = E_i(t)\braket{j}{i}.
\end{equation}
We can also consider the hamiltonian (with $\hbar=1$) acting on bare state to the left. For the state $\ket{g}$, for example, this leads to 
    \begin{align}
    \begin{split}
     \mel{g}{\hat{H}}{i} &= \begin{pmatrix}
        1 & 0 & 0 & 0
    \end{pmatrix}
    \frac{1}{2}\begin{pmatrix}
        0 & \Omega_p & 0 & 0 \\
        \Omega_p & 2\Delta & \Omega_s & \Omega_s \\
        0 & \Omega_s & 2\delta & 0 \\
        0 & \Omega_s & 0 & -2\delta\\
    \end{pmatrix}\ket{i}\\
    &=\frac{\Omega_p}{2}\braket{e}{i}. \label{eq:SupplPhase2}
     \end{split}
\end{align}
Combining the  equations for the states $\ket{g}$, $\ket{\uparrow}$, $\ket{\downarrow}$ with the relevant expressions  (eqns. \ref{eq:SupplPhase1}) allows us to express the elements of the four adiabatic eigenstates $\ket{a} \dotsi\ket{d}$ in the bare basis:
\begin{equation}\label{eq:SupplPhase3}
\begin{split}
    \braket{g}{i} = \frac{\Omega_p}{2E_i}\braket{e}{i}, \\
    \braket{\uparrow}{i} = \frac{1}{E_i - \delta}\frac{\Omega_s}{2}\braket{e}{i}, \\
    \braket{\downarrow}{i} = \frac{1}{E_i +\delta}\frac{\Omega_s}{2}\braket{e}{i}.
\end{split}
\end{equation}

The eigenenergies $E_i$ are given by the characteristic equation of the hamiltonian.
This is a quartic equation in $E_i$:
\begin{equation} 
  4E_i^4-f^2 E_i^2+\Omega_p^2\delta^2
    -4\Delta E_i\left(E_i^2-\delta^2\right)=0,
    \label{eq:quartic}
\end{equation}
with $f$ defined as 
\begin{equation}\label{appAfdef}
  f^2 = \Omega_p^2+2\Omega_s^2+4\delta^2.
\end{equation}

For $\Delta = 0$, the equation has these four roots
\begin{equation}\label{eq:SupplPhaseEigenenergies}
    E_i = \pm\frac{1}{2\sqrt{2}}
\sqrt{f^2\pm\sqrt{f^4-16\delta^2\Omega_p^2}}
  \end{equation}
which fully determine the eigenstates. The states $\ket{a}$ and $\ket{b}$ have eigenenergies closest to 0 
with $E_b = -E_a$. Expressed in the bare basis, via eqns.  \ref{eq:SupplPhase3}, these are the eigenstates given in eqns. \ref{eq:stateA} -- \ref{eq:ABEigenengergies} in the main text. 

The  eigenstates at the start of the STIRAP process are found by taking the delicate limit $\Omega_s \rightarrow 0$ whilst ensuring $\Omega_p/\Omega_s\rightarrow 0$. These are given by eqns. \ref{InitialState} and \ref{protostate} in the main text.

If $\delta$ is exactly zero, $\ket{g}$ is aligned fully with one of the adiabatic eigenstates. For  t$\rightarrow-\infty$ with $\delta \neq 0$, the ground state is the superposition 
\begin{equation}
    \ket{\psi(- \infty)} = \frac{1}{\sqrt{2}}\left(\ket{a(-\infty)} - \ket{b(-\infty)}\right) = \ket{g}.
\end{equation}

\section{Derivation of the relative phase} \label{sec:AppendixRelativePhase}
In the adiabatic limit, with zero geometric phase (see below), the state at time $t$ is given by eq.~\ref{protostate} in the main text.
This is simplified by factoring an overall phase factor and noticing that $E_b(t) = -E_a(t)$,
\begin{gather}
    \ket{\psi(t)} = \frac{1}{\sqrt{2}}\left(\ket{a(t)}-e^{i\Theta(t)}\ket{b(t)}\right)
    \label{eq:state}
\end{gather}
with $\Theta$ as
\begin{gather}
        \Theta(t) = 2\int_{-\infty} ^t E_a(t') dt'. \label{int_append}        % \label{eq: theta}
\end{gather}
This integral has the compact form
\begin{equation}
     \Theta(t) =\frac{-1}{\sqrt{2}}\int_{-\infty}^t\sqrt{f(t')^2-\sqrt{f(t')^4-g(t')^2}} dt',
\end{equation}
with the $f(t')$ defined in eq. \ref{appAfdef} and
    $g(t') = 4\delta\Omega_p$.
At very early times, when the Rabi frequencies are negligible, the integral evaluates to 0. However, during  the adiabatic transfer the Rabi frequencies are orders of magnitude larger than $\delta$. This means that  $f(t')^4\gg g(t')^2$, which permits us to expand the inner square-root. We set the start time $t=0$ to be late enough that $f\gg g$ but early enough no significant population transfer has occurred.  
\begin{equation}
    \Theta(t) \approx \frac{-1}{\sqrt{2}}\int_0^t \sqrt{\frac{g(t')^2}{2f(t')^2}} dt'.
\end{equation}
With $\delta$ much smaller than the Rabi frequencies,  the dynamic phase is
\begin{equation}
    \Theta(t) \approx -2\delta \int_0^t \frac{ dt'}{\sqrt{1+2(\Omega_s /\Omega_p)^2}}. 
    \label{eq: dynamicphase}
\end{equation}
Taking gaussian pulse envelopes leads to the final integral
\begin{equation}
    \Theta(t) = -2\delta\int_0^t\frac{dt'}{\sqrt{1+2r^2e^{-T^{-2}[(t'-\mu_s)^2-(t'-\mu_p)^2]}}}
    \label{eq:appendintegral}
\end{equation}
This integral has a closed form solution which leads to eq. \ref{solution}.

Using the expansion of the eigenstates in the bare basis, $\ket{\psi(t)}$ is
\begin{equation}
\begin{split}
    \ket{\psi(t)} &= \braket{e}{a}\left[\frac{\Omega_p}{2E_a}(1 + e^{i\Theta(t)})\ket{g}+(1 - e^{i\Theta(t)})\ket{e} \right. \\  &+\frac{\Omega_s}{2}\left(\frac{1}{E_a - \delta} + \frac{e^{i\Theta(t)}}{E_a + \delta}\right)\ket{\uparrow}  \\ &+\left.   \frac{\Omega_s}{2}\left(\frac{1}{E_a + \delta} + \frac{e^{i\Theta(t)}}{E_a - \delta}\right)\ket{\downarrow}\right].
\end{split}
\end{equation}

When $\delta$ is small 
$    e^{i\Theta(t)} \approx 1 + i\Theta(t),
$ and the relative phase between $\ket{\uparrow}$ and $\ket{\downarrow}$ becomes
\begin{multline}
    \phi(t) = \arg\left(2E_a + i(E_a-\delta)\Theta(t))\right) - \\ \arg\left(2E_a + i(E_a+\delta)\Theta(t))\right) .
\end{multline}
Since $\Theta$ is small  $\tan (\theta)\approx\theta$ which gives the simple result
$\phi(t) = -(\hbar\delta/{E_a})\Theta(t).$
Evaluating eqn. \ref{eq:appendintegral} gives    
\begin{equation}
    \begin{split}
        \phi(t)& = \frac{2\delta^2 T^2}{E_a(t) (\mu_p-\mu_s)}\\ &\times\left(\arctanh(\sqrt{1+2r^2\exp(\tfrac{\mu_p-\mu_s}{T^2}(\mu_p +\mu_s-2t))} \right.\\
        &-\arctanh\left.(\sqrt{1+2r^2\exp(\tfrac{\mu_p-\mu_s}{T^2}(\mu_p +\mu_s))}\right),
        \label{solution}
    \end{split}
    \end{equation}
 with $r = \Omega_{0s}/\Omega_{0p}$.   
 Defining dimensionless variables
\begin{equation}
\begin{split}
    \tau &= \frac{2(\mu_p -\mu_s)}{T^2}t ,\\
    \epsilon (\tau) &= \frac{\delta^2T^2 }{(\mu_p-\mu_s)E_a(\tau) }, \\
    \gamma &= 2r^2e^{\frac{\mu_p^2-\mu_s^2}{T^2}} \label{77},
 \end{split}
 \end{equation}
 gives
 \begin{equation}
    \phi(\tau) = \epsilon (\tau)\ln{\left[\frac{1-\sqrt{1+\gamma}}{1+\sqrt{1+\gamma}}\frac{1+\sqrt{1+\gamma e^{-\tau}}}{1-\sqrt{1+\gamma e^{-\tau}}}\right].} \label{formula}
 \end{equation}

The discussion above has assumed $\Delta =0$. We can relax this assumption by treating $\Delta$ as a perturbation: $\delta\hat{H} = \Delta \ket{e}\bra{e}$. Therefore, the perturbations on the energies are $\bra{a}\delta\hat{H}\ket{a} = \bra{b}\delta\hat{H}\ket{b} = \Delta |N|^2$ from eqns. (3) -- (5). Since the extra phase contribution of the perturbed energies in eq. (10) is the same for both $\ket{a}$ and $\ket{b}$, the total contribution is zero to first order. Numeric calculations confirm that the contribution of $\Delta$ to the phase is small and is associated with non-adiabatic processes.  
\vspace{1ex}
\section{Geometric Phase}\label{sec:AppendixGeometricPhase}
In the adiabatic limit, a wavefunction acquires an additional phase
\begin{equation}
    \gamma = i\int_0^t \bra{\psi(t')}\frac {\partial}{\partial t'}\psi(t') \rangle dt'.
\end{equation}
For a state $\ket{\psi}$ that is purely real $\gamma=0$.
For the STIRAP process, we are interested in the relative phase between $\ket{a}$ and $\ket{b}$:
\begin{equation}
    \gamma_{ab} = i\int_0^t\braket{b(t')}{\dot{b(t')}}-i\int_0^t\braket{a(t')}{\dot{a(t')}}dt'.
\end{equation}
For  the hamiltonian in eq. \ref{hamiltonian}, the eigenstates are real. However, a phase difference between the Stokes and Pump lasers introduces complex components through the phase difference in the couplings. The hamiltonian
\begin{align}
    \begin{split}
     \Tilde{H}= 
    \frac{1}{2}\begin{pmatrix}
        0 & \Omega_p & 0 & 0 \\
        \Omega_p & 2\Delta & \Omega_se^{-i\varphi_1} & \Omega_se^{-i\varphi_2} \\
        0 & \Omega_se^{i\varphi_1} & 2\delta & 0 \\
        0 & \Omega_s e^{i\varphi_2}& 0 & -2\delta\\
    \end{pmatrix}\label{eq:SupplHam2}
     \end{split}
\end{align}
now has complex eigenstates, where the time dependence is carried in the real coefficients $\Omega_s$, $\Omega_p$. It is easy to show that the eigenstates of this hamiltonian yield inner products $\bra{a}\dot{a}\rangle$ and $\bra{b}\dot{b}\rangle$ that are equal, so $\gamma_{ab}=0$ even if the individual states acquire a geometric phase. 

\section{Reverse STIRAP} 
\label{sec:reversestirap}
\begin{figure}[h!]
    \centering
    \vspace{0.1em}
    \includegraphics[scale=0.6]{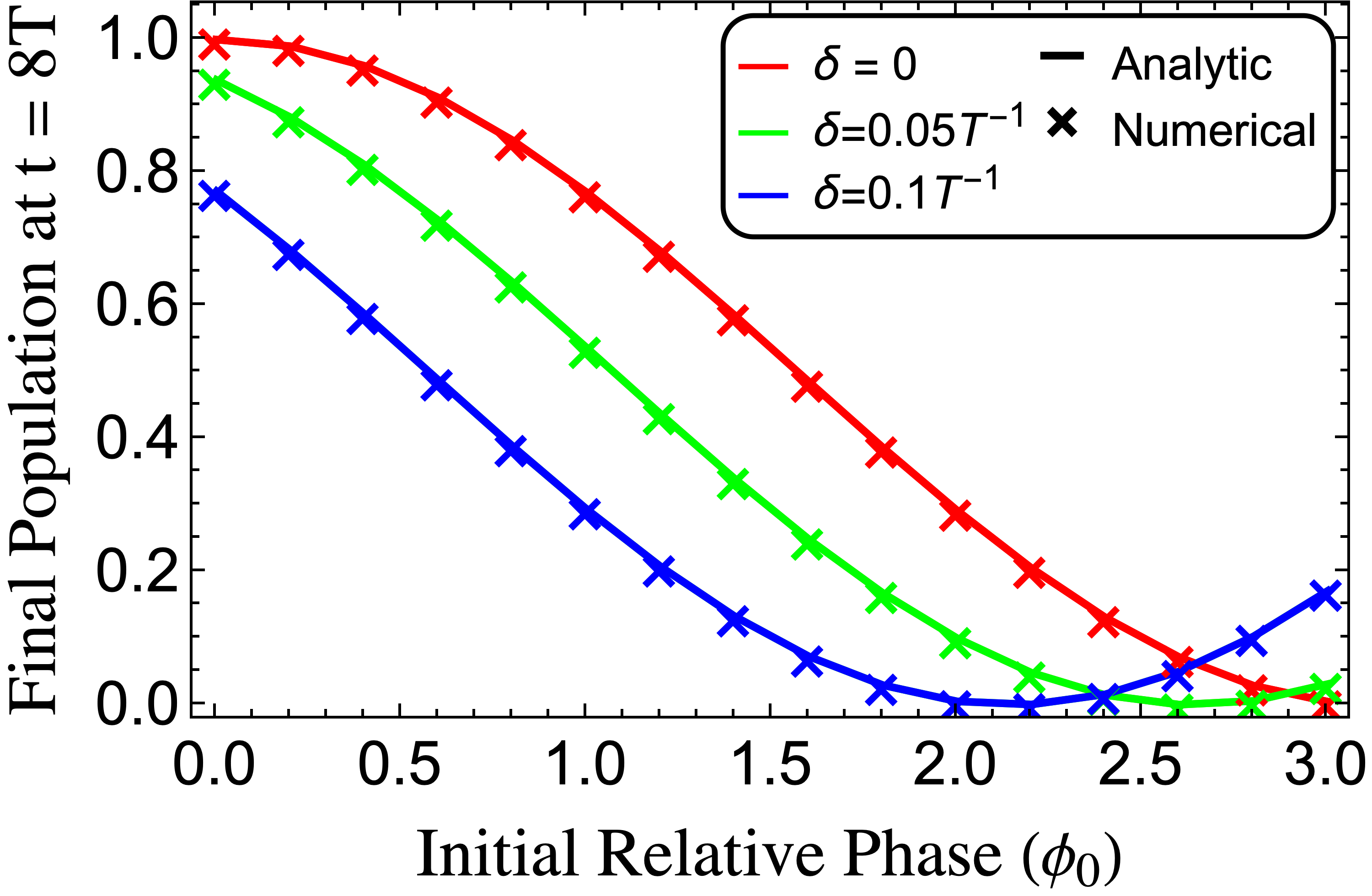}
    \caption{Population transfer to the ground state for reverse STIRAP versus the relative phase of the incoming state. The solid lines show numerical results while crosses are analytic. }
    \label{fig:reversestirap}
\end{figure}

In the time reversal violation experiment, the initial state is first transferred to a superposition, then transferred back after an evolution time.
This reverse STIRAP is the time-reversed version of the process we have considered. The initial state is a superposition state, and the order of the Stokes and pump beams is reversed. A superposition state with an arbitrary relative phase is
 
\begin{equation}
    \frac{1}{\sqrt{2}}(\ket{\uparrow}+e^{i\phi_0}\ket{\downarrow})=\frac{1}{\sqrt{2}}(\ket{a(-\infty)}-e^{i\phi_0}\ket{b(-\infty)}).
\end{equation}
In the adiabatic approximation
this evolves at time $t$ to
\begin{equation}
    \ket{\psi(t)}=\frac{1}{\sqrt{2}}(\ket{a}-e^{i(\Theta (t)+\phi_0)}\ket{b}),
\end{equation}
where $\Theta (t)$ is defined in equation \ref{int_append}. The final-state population is then
\begin{multline}
    \abs{\braket{g}{\psi(t)}}^2= \frac{1}{2}\abs{\braket{g}{a}-e^{i(\Theta (t)+\phi_0)}\braket{g}{b}}^2 \\ = \frac{\abs{N}^2\Omega_p^2}{2E_a^2}\cos^2\left(\frac{\phi_0-\Theta (t)}{2}\right).
    \label{eq:groundstateprediction}
\end{multline}
Equation \ref{eq:groundstateprediction} describes how the final ground state population depends on the relative phase ($\phi_0$) of the incoming superposition state. For $\delta>0$, the phase of the incoming state continues to evolve \textit{during} STIRAP with a `trivial' part due to $\delta$ and potentially a part due to STIRAP. 
The trivial phase at time time $t_{end}$ is just $2\delta \cdot t_{end}$. We find numerically when $t_{end} = 5T$ (when nearly all the population has transferred) gives 
\begin{equation}
    -\Theta(\infty) \approx 2\delta \cdot t_{end}.
    \label{eq:spintheta}
\end{equation}
to within 1\%. There is no significant phase other than the trivial one. Figure \ref{fig:reversestirap} compares numerical results to formula  \ref{eq:groundstateprediction}. The adiabatic approximation is very good for the reverse process.

\newpage
\bibliography{bibliography,references}
\end{document}